\documentclass{article}

\input{tcilatex}
\begin{document}

\title{\textbf{Background dielectric permittivity: material constant or
fitting parameter?}}
\author{A.P. Levanyuk,$^1$ B.A. Strukov,$^2$ and A. Cano$^3$ \\
{\small $^1$University of Washington, Department of Physics, Seattle, WA
98105 USA} \\
{\small $^2$Moscow State University, Moscow.119991, Russia }\\
{\small $^3$CNRS, University Bordeaux, ICMCB, UPR 9048, F-33600 Pessac,
France} }
\maketitle

\begin{abstract}
The concept of background dielectric constant proposed by Tagantsev,
together with the distinction between critical and non-critical electric
polarizations as a natural extension for the order parameter of
ferroelectric phase transitions, is critically discussed. It is argued and
exemplified that, even if these quantities can be very useful for obtaining
qualitative and semi-quantitative results from phenomenological modeling,
they cannot be introduced in a self-consistent manner.
\end{abstract}

\section{Introduction}

The so-called depolarizing field is at the heart of many specific properties
of ferroelectric materials. This field is due to (longitudinal) spatial
variations of the polarization itself, rather than due to an external
source. The key role of the depolarizing field in the formation of
ferroelectric domains, for example, was already discussed in the early
papers on ferroelectrics (see e.g. \cite{Mitsui,Kanzig1}) and thanks to a
number of complementary studies is rather well understood at present (see
e.g. \cite{junquera08,bratkovsky08}). The impact of the depolarizing field
on the fluctuations of the polarization and its relaxation to equilibrium
has also been studied by many authors (see e.g. \cite{Krivoglaz}). In this
respect, one of the most spectacular effects experimentally verifiably was
obtained for the relaxation time of the inhomogenous distribution of
polarization in uniaxial ferroelectrics. This distribution can be created by
means an acoustic wave, and the attenuation coefficient was predicted to
depend on the propagation direction of this wave \cite{Geguzina}. Such a
dependence was subsequently observed in TGS, but with a dramatic
quantitative disagreement with the theoretical prediction \cite{Minaeva}.
This disagreement was interestingly explained by Tagantsev \cite{Tag86} by
separating the total polarization into \textquotedblleft order-parameter"
(ferroelectric) and \textquotedblleft background" (non-ferroelectric)
contributions to the total polarization, and further characterizing the
latter by means of a \textquotedblleft background'' (or ``base") dielectric
constant. Thus, the difference between theory and experiment was resolved
within a Landau-like framework by fitting the value of the background
dielectric constant.

In view of this remarkable success, the development of a Landau theory in
which the status of the background dielectric constant is upgraded from mere
fitting parameter to true material constant has an obvious appeal. This
development has been advocated in \cite{Tag08} by invoking some general
physical arguments. At first glance, these arguments seem rather natural and
convincing. It is clear that, when it comes to its symmetry properties, the
order parameter of a proper ferroelectric transition is equivalent to the
electric polarization. At the same time, it is also clear that the total
polarization generally contains multiple contributions either ionic,
electronic, or both, and that only one specific pattern of them can be
identified to the actual order parameter (i.e. only one specific pattern is
behind the ferroelectric instability). In consequence, together with the
\textquotedblleft critical\textquotedblright\ polarization, there are many
\textquotedblleft non-critical\textquotedblright\ polarizations that respond
to the presence of electric fields but do not emerge spontaneously right
after the phase transition. According to \cite{Tag08}, the background
dielectric constant could be considered as a new material parameter
characterizing the non-critical contributions to the total polarization.
However, as we show below, this viewpoint has strong conceptual limitations.
The Landau theory is in fact quantitatively correct for describing
asymptotic behaviors at second-order phase transitions (with the
reservations mentioned in Ref. \cite{Comment1}). We shall argue that the
incorporation of a non-ferroelectric 
polarization spoils the internal consistency of the Landau theory of phase
transitions and, in consequence, makes it impossible to quantify the
background dielectric constant from experimental measurements or
first-principle calculations.

This limitation does not mean that the theoretical framework proposed in 
\cite{Tag08} is impractical. In fact, there are cases in which the
non-ferroelectric contributions to the total polarization play an important
role. The failure is simply that these contributions cannot be incorporated
in a self-consistent way\textbf{.} 
In other words, Tagantsev's approach cannot be considered as a rightful
extension of the Landau theory of phase transitions, but rather as a
Landau-like phenomenological modeling 
providing qualitative or semi-quantitative results only. Beyond that, we
recognize that a very important merit of \cite{Tag08} is to stimulate the
discussion about the physical meaning of order parameter for proper
ferroelectrics. This conceptual discussion is crucial, not only to clarify
and to advance the state-of-the-art in phenomenological modeling, but also
to build bridges with complementary approaches, notably
first-principle-based descriptions. This paper is expected to contribute
constructively to this discussion.

The paper is organized as follows. In Sec.2 we introduce the background
dielectric constant proposed in \cite{Tag86} by analyzing the depolarizing
field created in a ferroelectric slab. In Sec.3 we critically analyze a
Landau-like free energy proposed in Ref.\cite{Tag08}. In Sec.4 we discuss
the limits of the Landau theory \cite{Comment1} using a fairly general
phenomenological model. In Sec.5 we summarize the conceptual results of the
paper.

\section{Preliminaries}

The concept of background dielectric constant can be easily understood by
considering the depolarizing field in a slab of an uniaxial ferroelectric.

Consider the simplest case of homogeneous out-of-plane polarization, $%
\mathbf{P}=P\hat{\mathbf{z}}$, in which the depolarizing field is generated
by the corresponding bound charges at the interfaces. This field can be
obtained from the electrostatic equation $\nabla \cdot \mathbf{D}=0$ \label%
{1} that must be satisfied inside the ferroelectric. Here $\mathbf{D}%
=\varepsilon _{0}\mathbf{E}+\mathbf{P}$ is the electric displacement vector,
where $\mathbf{E}=\mathbf{E}_{\mathrm{dep}}$ in our case. The symmetry of
the problem is such that $D_{x}=D_{y}=0$ and $D_{z}=0$ due to the boundary
conditions. Consequently, the depolarizing field is \ 
\begin{equation}
\mathbf{E}_{\mathrm{dep}}=-{\frac{1}{\varepsilon _{0}}}\mathbf{P}.  \label{2}
\end{equation}%
The key question raised in \cite{Tag86} can be reformulated in the following
way: Can $\mathbf{E}_{\mathrm{dep}}$ be computed within the Landau theory of
phase transitions, i.e., from the corresponding order parameter?

The order parameter represents the variable with respect to which the
paraelectric phase losses its stability in a second-order transition. In
view that it is polar by symmetry, 
Ginzburg originally identified this variable with the total polarization
without much additional justification \cite{Ginzburg45}. It turns out that,
sufficiently close to the paraelectric-ferroelectric transition this
identification is always possible even though in any real system there are
multiple partial polarizations (due to different ions and contributions of
electronic clouds) that, from the symmetry point of view\textbf{,} are
equally good to be the order parameter. Physically, the order parameter
corresponds to a specific linear combination of partial polarizations.
However, the knowledge of this linear combination is not necessary for the
Landau theory where this is not implied. In any case, the order parameter is
only one of the many variables characterizing the polar microstructure of
the system, and all these variables contribute to the total polarization. 
The correctness of Ginzburg's\textbf{\ }identification relies in the
ferroelectric instability itself: the instability implies a very large
response of the order-parameter variable to infinitesimally small stimuli,
i.e. softness, such that the total polarization in an external electric
field is dominated by this \textquotedblleft critical\textquotedblright\
contribution in the 
vicinity of the instability point (for an explicit treatment see Sec. 4.1
below). Similarly, the change of the polar microstructure in a
short-circuited sample is also dominated by the change in the order
parameter close to the instability. However, there are problems in which the
rest of polar variables cannot be neglected as their influence can be rather%
\textbf{\ }strong. Let\textbf{\ }us see if this is the case in the the above
problem of a ferroelectric slab with uniform polarization. For this, let us
write the total polarization as\textbf{\ $\mathbf{P}=\mathbf{P}_{f}+\mathbf{P%
}_{r}$ }where\textbf{\ $\mathbf{P}_{f}$ }represents the ferroelectric order
parameter\textbf{\ }and\textbf{\ $\mathbf{P}_{r}$ }encloses the rest of
contributions to the total polarization\textbf{.} In this qualitative and
preliminary consideration it is natural to assume that $\mathbf{P}_{r}$ is
analogous to the regular polarization of a standard (non-ferroelectric)
dielectric. In particular, its dependence on the electric field is proposed
to be $\mathbf{P}_{r}=\varepsilon _{0}(\varepsilon _{b}-1)\mathbf{E}$ where $%
\varepsilon _{b}$ is the background dielectric constant \cite{Tag08}. 
Substituting this into Eq. (\ref{2}) we obtain: 
\begin{equation}
\mathbf{E}_{\mathrm{dep}}=-{\frac{1}{\varepsilon _{0}\varepsilon _{b}}}%
\mathbf{P}_{f}.  \label{3}
\end{equation}%
If $\varepsilon _{b}$ is analogous to dielectric constant of a standard
dielectric, its typical value may be several tens. This means that when the
depolarizing field is expressed in terms of the order parameter, a strong
influence of the\textbf{\ }regular degrees of freedom can indeed be
expected. 

Since this situation can be encountered in this and in a number of related
problems, the following questions are quite reasonable. Is it possible to
develop a theory in which the regular degrees of freedom are included in a
consistent way\textbf{?} 
Can the above scheme be more than qualitative, \textbf{with} the background
dielectric constant \textbf{being} more than a vaguely defined fitting
parameter? These are the questions that we try to answer below. 

\section{General approach}

The description of the ferroelectric transition proposed in Ref. \cite{Tag08}
implements the above division between critical and non-critical
contributions to the total polarization. Specifically, the standard Landau
free energy for an (unixaxial) ferroelectric is replaced by 
\begin{equation}
F=\frac{\alpha }{2}P_{f}^{2}+\frac{\beta }{4}P_{f}^{4}+\frac{1}{2
\varepsilon _{0}\chi _{b}}P_{r}^{2}-\left( P_{f}+P_{r}\right) E,  \label{4}
\end{equation}%
where $\chi _{b} = \varepsilon_{b} - 1$. Let us check the internal
consistency of the proposed framework by working out different examples.

\subsection{Temperature dependence of the susceptibility}

Let us first consider the linear response of the system in the paraelectric
phase. In the case of a temperature-induced second-order transition, it is
customary to take $\alpha =\alpha ^{\prime }\left( T-T_{c}\right) $ with $%
T_{c}$ being the transition temperature. According to Eq. (\ref{4}) we then
have 
\begin{equation}
P_{f}=\frac{1}{\alpha ^{\prime }(T-T_{c})}E,  \label{5}
\end{equation}%
\begin{equation}
P_{r}=\varepsilon _{0}\chi _{b}E,  \label{6}
\end{equation}%
and hence 
\begin{equation}
\chi \equiv {\frac{\partial P}{\partial E}}=\varepsilon _{0}\left( \frac{C}{%
T-T_{c}}+\chi _{b}\right)   \label{5a}
\end{equation}%
where $C=1/(\varepsilon _{0}\alpha ^{\prime })$. As we see, the Curie-Weiss
behavior is obtained from $P_{f}$ while the temperature-independent response
is due to $P_{r}$. At first glance, this seems to be a clear-cut way of
determining these two polarizations (and hence of measuring the background
permittivity). However, if one considers the next order term in $\alpha
=\alpha ^{\prime }\left( T-T_{c}\right) +\alpha ^{\prime \prime }\left(
T-T_{c}\right) ^{2}$, the critical polarization becomes 
\begin{equation}
P_{f}=\frac{E}{\alpha ^{\prime }\left( T-T_{c}\right) +\alpha ^{\prime
\prime }\left( T-T_{c}\right) ^{2}}=\left( \frac{1}{\alpha ^{\prime }\left(
T-T_{c}\right) }-\frac{\alpha ^{\prime \prime }}{\alpha ^{\prime 2}}+%
\mathcal{O}\big((T-T_{c})\big)\right) E,  \label{7}
\end{equation}%
and consequently 
\begin{equation}
\chi =\varepsilon _{0}\left( \frac{C}{T-T_{c}}+\widetilde{\chi }_{b}+%
\mathcal{O}\big((T-T_{c})\big)\right) ,  \label{7a}
\end{equation}%
where $\widetilde{\chi }_{b}=\chi _{b}-{\frac{\alpha ^{\prime \prime }}{%
\varepsilon _{0}\alpha ^{\prime 2}}}$. This means that, strictly speaking,
the temperature-independent part of the total response contains
contributions from both $P_{f}$ and $P_{r}$ and, consequently, its
connection to the background dielectric constant is ambiguous. Or, put in
more constructive terms, the identification of the background dielectric
constant with the temperature-independent part of the electric
susceptibility is subjected to the condition that the $\mathcal{O}\big(%
(T-T_{c})^{2}\big)$ terms in $\alpha $ can be neglected.

\subsection{Nonlinear couplings \& susceptibility}

Another shortcoming is that $P_r$ contains corrections due to the presence
of high-order terms that are eventually indistinguishable from the obtained
from $\varepsilon_b $ itself in Eq. (\ref{6}). In fact the quantities $P_{f}$
and $P_{r}$ have the same transformation properties, and therefore the
generalized free energy (\ref{4}) can contain additional coupling terms. At $%
T=T_c$ the bi-linear coupling $P_{f} P_{r}$ is absent as argued in \cite%
{Tag08} and we will demonstrate below. However, there is no reason to expect
the absence of high-order terms even at $T=T_c$. So let us consider these
extra terms and address the following question: is it possible to single out
critical and non-critical polarizations by studying the electric-field
response at $T=T_{c}$?

Consider first the coupling $\gamma P_{r}P_{f}^{3}.$ Including this coupling
into the free energy Eq. (\ref{4}) and further minimizing with respect to $%
P_{f}$ and $P_{r}$ we obtain 
\begin{eqnarray}
\beta P_{f}^{3}+3\gamma P_{r}P_{f}^{2} &=&E,  \label{8} \\
\frac{1}{\varepsilon _{0} \chi_b}P_{r}+\gamma P_{f}^{3} &=&E.  \label{9}
\end{eqnarray}%
If the external field is sufficiently low this givies%
\begin{equation}
P_{f}\approx \left( E/\beta \right) ^{1/3},  \label{10.1}
\end{equation}
\begin{equation}
P_{r}\approx \left( 1-\frac{\gamma }{\beta }\right) \varepsilon
_{0}\chi_{b}E.  \label{10}
\end{equation}
By comparing Eqs. (\ref{10}) and (\ref{6}) we can see that the high-order
coupling in fact generates a correction to $P_r$ that is also linear in $E $%
. Should we take it into account? If yes, then not only it. In fact,
combining Eqs. (\ref{8}) and (\ref{9}) one obtains 
\begin{equation}
\beta P_{f}^{3}-3\gamma ^{2} \varepsilon _{0}\chi_{b}P_{f}^{5}=\left(
1-3\gamma \varepsilon _{0}\chi_{b}P_{f}^{2}\right) E.  \label{11}
\end{equation}%
In this equation there appears a $P_{f}^{5}$ term that plays the same role
as the obtained from a sixth-order $P_{f}^{6}$ term in the generalized free
energy. This means that, if we include $P_{r}$ in Eq. (\ref{4}), then we
have to do it by introducing not only the $P_{r}^2$ term in the initial free
energy, but also the $P_{f}^{3}P_{r}$ and $P_{f}^{6}$ terms that essentially
play the same role. Or in other words, the field dependence of $P_{f}$ and $%
P_{r}$ turns out to be determined not only by the background dielectric
constant $\varepsilon_b$, but also by the coefficients of these high-order
terms. The incorporation of all these terms does not sound very practical,
and the extended framework thus losses its appeal.


We recall that the internal consistency of the Landau theory lies on the
fact that only asymptotic behaviors near the transition point are
considered. In this sense, Eq. (\ref{10.1}) provides such an asymptotic
behavior while Eq. (\ref{10}) is beyond the scope of the Landau theory since 
${E\to 0}\mathrm{lim}{P_{r}/P_{f}}= 0$.

\subsection{Spontaneous polarization}

Another example is temperature dependence of the total spontaneous
polarization ($E=0$). Can we individuate its critical and non-critical
contributions? According to Eq. (\ref{4}) $P_{r}=0$ at zero field, but
taking into account the coupling $\gamma P_{r}P_{f}^{3}$ one obtains $%
P_{r}\propto P_{f}^{3}\propto (T-T_{c})^{3/2}$. However, the same dependence
is obtained from the bilinear coupling which, as we will see the next
section, has the form $\alpha _{12}\left( T-T_{c}\right) P_{f}P_{r}$. In
addition, the terms $\mathcal{O}\big((T-T_{c})^{2}\big)$ and $\mathcal{O}%
(T-T_{c})$ in the Taylor series expansion of the coefficients $\alpha $ and $%
\beta $ respectively also result into a similar\textbf{\ $(T-T_{c})^{3/2}$ }%
dependence of\textbf{\ $P_{f}$}. Consequently, we see once again that $P_{r}$
and $P_{f}$ are poorly defined quantities with no direct experimental
access: even if the subdominant $\left( T-T_{c}\right) ^{3/2}$ contribution
to the total spontaneous polarization is determined experimentally, one
cannot conclusively say if this is due to the lowest-order contribution to $%
P_{r}$, high-order contributions to $P_{f}$, or both.

\subsection{Inhomogeneous depolarizing fields}

The notion of background dielectric constant plays its most important role
in problems related to\emph{\ }homogeneous or almost homogeneous
depolarizing field. There are many problems, however, where\emph{\ }this%
\emph{\ }field is strongly inhomogeneous. A typical example is formation of
domain structure in not extremely thin films. Here the scale of changes in
the depolarizing field is much less than the film thickness. problem of the
. In this case the response of the system perpendicular to the ferroelectric
axis comes to the scene on top of the non-critical response along the
ferroelectric direction. This is because an inhomogenous depolarizing field
necessarily has to have multiple components according to the electrostatic
equation $\nabla \times \mathbf{E}=0$. As a result, $\varepsilon _{b}$ is
generally in competition with well-defined material constants such as the
dielectric constant perpendicular to the ferroelectric axis. The good news
is that, even if the results obtained from the Landau-like formalism
nominally depend on $\varepsilon _{b}$, they can nonetheless be dominated by
true material constants and hence be robust results. An example of this
situation which \ can be found in Ref.\cite{Burc} where the Landau-like
approach was used to interpret\emph{\ }\ experimental data of Ref.\cite%
{Tenne}. Strong effect of film thickness ($l$) on ferroelectric phase
transition temperature ($T_{c}$) in non-electroded BaTiO$_{3}$ films.on SrTiO%
$_{3}$ substrate was revealed in this paper: at changing the film thickness
from $1.6\unit{nm}$ to $10\unit{nm}$ the phase transition temperature
changed from $70\unit{K}$ to $925\unit{K}.$Theoretically this phase
transition is expected to be into multidomain state .The only parameter of
the Landau-like approach unknown from independent experiments was $%
\varepsilon _{b}$. It proved out that changing of $\varepsilon _{b}$ from $1$
to $10$ is almost unnoticeable for the theoretical $T_{c}\left( l\right) $
curve while its changing from $10$ to $100$ has small but still noticeable
effect for $l<5\unit{nm}.$

\section{Model approach}

In order to further clarify the physical meaning of the variables introduced
in \cite{Tag08}, it is instructive to reconsider the loss of stability of
the paraelectric phase from the perspective of a specific model and its
effective Hamiltonian (or incomplete free energy). For this, let us consider
the simplest situation in which we have just two ``microscopic''
contributions to the total polarization, $P = P_1 + P_2$, and the most
general effective Hamiltonian allowed by symmetry: 
\begin{equation}
F\left( P_{1},P_{2};T,E\right) =\frac{a _{1}}{2}P_{1}^{2}+\frac{a _{2}}{2}%
P_{2}^{2}+a_{12} P_{1}P_{2}+...-\left( P_{1}+P_{2}\right)E .  \label{12}
\end{equation}
The individual polarizations $P_1$ and $P_2$ can be, for example, the ionic
polarizations of two different atoms or the ionic and electronic
polarizations of the same atom. For the sake of concreteness we consider the
temperature as the control parameter. Accordingly, the coefficients of this
Hamiltonian are assumed to be (unknown) functions of this parameter.

\subsection{Individual vs. total polarization}

The loss of stability of the paraelectric phase can be analyzed from the
linearized equations of state (or equilibrium equations):%
\begin{equation}
a _{1}P_{1}+a_{12} P_{2}=E ,  \label{13}
\end{equation}%
\begin{equation}
a _{2}P_{2}+a_{12} P_{1}=E .  \label{14}
\end{equation}%
The stability of the paraelectric phase ($P_1=P_2=0$) as the ground state of
the system requires $a _{1}$, $a _{2}$, and $a _{1}a_{2}-a_{12}^{2}$ to be $%
>0$. Conversely, the paraelectric phase losses its stability at the point at
which $a_{1}a_{2} - a_{12} ^{2} =0 $. Since this point can be reached from
the paralectric phase where $a_{1}$ and $a_{2}$ are $>0$, this means that,
in general, both $a_{1}$ and $a _{2}$ are positive at the transition point
(unless $a_{12} = 0$). In addition, the sign of $a_{12}$ determines the
precise state that emerges in the transition: $P_1$ and $P_2$ have the same
sign (parallel) if $a_{12}<0$, while they have opposite sign (anti-parallel)
if $a_{12}>0$. In the following we consider $a_{12}<0$, although essentially
the same is obtained for $a_{12}<0$.

Near the transition point it is generically possible to define the Taylor
expansion $a_{1}a_{2}-a_{12}^{2}\equiv \Delta =\Delta ^{\prime
}(T-T_{c})+\dots $, where $T_{c}$ is the transition temperature and $\Delta
^{\prime }$ a positive constant (such that $a_{1}a_{2}-a_{12}^{2}>0$ for $%
T>T_{c}$). In the paraelectric phase, this automatically gives the
Curie-Weiss-law behavior for both individual and total polarizations: 
\begin{equation}
P_{1(2)}=\chi _{1(2)}E,  \label{15}
\end{equation}%
\begin{equation}
P=P_{1}+P_{2}=\chi E,  \label{16}
\end{equation}%
where, according to Eqs. (\ref{13}) and (\ref{14}), 
\begin{equation}
\chi _{1(2)}\approx \frac{C_{1(2)}}{T-T_{c}},
\end{equation}%
\begin{equation}
\chi \approx \frac{C}{T-T_{c}},  \label{16a}
\end{equation}%
with $C_{1(2)}=[a_{2(1)}\left( T_{c}\right) -a_{12}\left( T_{c}\right)
]/\Delta ^{\prime }$ and $C=[a_{1}\left( T_{c}\right) +a_{2}\left(
T_{c}\right) -2a_{12}\left( T_{c}\right) ]/\Delta ^{\prime }$. Note that all
these susceptibilities are positive since $a_{1(2)}(T_{c})>0$ and $%
a_{12}(T_{c})<0$.

As we see, all these polarizations display the same divergent behavior close
the transition point and, consequently, all of them can be considered as
critical (or, conversely, none of them is non-critical). Accordingly, the
order parameter of the transition can be associated to either of them. In
this sense, the total polarization is a rather natural choice as Ginzburg
originally considered in \cite{Ginzburg45}. However, to obtain Ginzburg's
free energy, one still has to minimize (or integrate out) over all the
variables that do not contribute to the total polarization. In the case of
the effective Hamiltonian (\ref{12}), this can be done by changing to the
variables 
\begin{equation}
P = P_1+ P_2,
\end{equation}
\begin{equation}
Q = P_1- P_2.
\end{equation}
Thus, in terms of these variables the Hamiltonian reads 
\begin{equation}
F\left( P,Q;T,E\right) =\frac{a_{1}+a _{2}+2a_{12} }{8}P^{2}+\frac{a_{1}+a
_{2}-2a_{12}}{8}Q^{2}+\frac{a_{1}-a_{2}}{4}PQ+...-PE  \label{17}
\end{equation}%
and the minimization over $Q$ yields 
\begin{equation}
F\left( P,Q\left( P\right) ;T,E\right) =\frac{1}{2 \chi}P^{2}+...-PE.
\label{18}
\end{equation}%
The functional obtained in this way corresponds to the standard free energy
in the Landau theory of phase transitions.

\subsection{Critical and non-critical polarizations}

In order to individuate critical and non-critical polarizations in our model
(and hence the background dielectric constant) we have to proceed
differently. The key point is to identify the actual degree of freedom (or
generalized coordinate) with respect to which the paraelectric phase becomes
unstable. That is, the degree of freedom describing the structural changes
that will emerge spontaneously right after the phase transition.
Mathematically, this corresponds to the eigenvector of the quadratic form in
Eq. (\ref{12}) whose eigenvalue vanishes and hence defines the transition
point. By performing standard linear algebra, one can easily find that the
linear combinations 
\begin{equation}
P_{f} = {\frac{1-c}{1+c^2 }} (P_1 - c P_2),
\end{equation}%
\begin{equation}
P_{r} = {\frac{1+c }{1+c^2 }} (cP_1 + P_2),
\end{equation}%
with $c =\frac{a _{1}\left( T_{c}\right) }{a_{12} \left( T_{c}\right) }=%
\frac{a_{12} \left( T_{c}\right) }{a_{2}\left( T_{c}\right) }$, represent
the eigenvectors of the model at $T_c$. Accordingly, the effective
Hamiltonian can be written as 
\begin{equation}
F\left( P_{f},P_{r};T,E\right) =\frac{A_{1}^{\prime }\left( T-T_{c}\right) }{%
2}P_{f}^{2}+\frac{A_{2}}{2}P_{r}^{2}+A_{3}^{\prime }\left( T-T_{c}\right)
P_{f}P_{r}+...-\left( P_{f}+P_{r}\right) E,  \label{22}
\end{equation}%
where 
\begin{equation}
A_{1}^{\prime }=\frac{a _{1}^{\prime }(T_{c})a _{2}(T_{c})+a_{2}^{\prime
}(T_{c})a _{1}( T_{c}) -2a_{12}^{\prime }( T_{c})a_{12}( T_{c}) }{a_{2}(
T_{c}) \left( 1-c \right) ^{2}},  \label{23}
\end{equation}%
\begin{equation}
A_{2}=\frac{a_{1}^{2}( T_{c}) +a_{2}^{2}( T_{c}) +2a_{12} ^{2}( T_{c}) }{%
a_{2}( T_{c}) \left( 1+c\right) ^{2}},  \label{24}
\end{equation}%
\begin{equation}
A_{3}^{\prime }=a_{12}^{\prime }( T_{c})+\frac{a_{1}^{\prime }( T_{c})-a
_{2}^{\prime }( T_{c})}{1-c^{2}}c.  \label{25}
\end{equation}
to the lowest relevant order in $T-T_c$. In this model, $P_f$ is therefore
the critical polarization while $P_r$ is the non-critical one. In fact, the
effective Hamiltonian can be related to Eq. (\ref{4}) by identifying 
\begin{equation}
A_{1}^{\prime }\left( T-T_{c}\right) \leftrightarrow \alpha , \qquad A_{2}
\leftrightarrow{\frac{1}{\varepsilon _{0}\chi_b}}.
\end{equation}
Compared to Eq. (\ref{4}), however, Eq. (\ref{22}) contains an additional
coupling term $P_{f}P_{r}$. This coupling vanishes at $T=T_{c}$ since, by
construction, only $P_{f}$ is behind the stability. At $T\not=T_{c}$, in
contrast, the coupling is nonzero, which simply tells us that the physical
content of the variables diagonalizing the Hamiltonian is different at
different temperatures. In Ref. \cite{Tag08}, however, this coupling was put
to zero also for $T\not =T_{c}$, which does not have an obvious
justification.

We find it instructive to discuss the possible origin of this subtle
mistake. In fact, we feel that two of the authors might have been reponsible
for it, as they made the same before Ref. \cite{Tag08} and in Refs. \cite%
{Strukov},\cite{Levanyuk}. It traces back to the tacit assumption that the
order parameter can be directly identified with a microscopic variable.
Strictly speaking, the Landau free energy is formally obtained by i)
identifying the complete set of microscopic variables with the same
transformation properties ii) integrating out over the rest, iii) choosing
the appropriate linear combination of the remaining variables and iv)
minimizing over the rest. The choice in step iii) is determined by physical
considerations related to the specific problem under consideration. And in
step iv) integration can be replaced by a simple minimization provided that
the corresponding degrees of freedom have measure zero compared to the total
degrees of freedom. But, beyond this, one essential point to be kept in mind
that, from step ii), one is dealing with an effective Hamiltonian whose
parameters, unlike in the initial Hamiltonian, depend on the corresponding
control parameters (e.g. temperature or pressure). In consequence, its
diagonalization is in general different for different values of the these
control parameters. As we have seen, the latter plays a role when critical
and non-critical order parameters are considered beyond the standard Landau
theory of phase transitions.

\section{Conclusions}

We have illustrated that the generality, consistency, and the
model-independency of the Landau theory of phase transitions is not
automatically inherited by Landau-like phenomenological models, even when
they logically seek to extend the perimeter of action of the Landau theory.
By considering the ferroelectric case, we have demonstrated that the
incorporation of non-critical polarizations and the corresponding background
dielectric constant is possible, but at the expense of sacrificing the full
internal consistency of the theory. The background dielectric constant, in
particular, turns out to be a mere fitting parameter --and not a true
material constant in its own right. This parameter still retains a more or
less precise physical meaning. In consequence, some physical intuition is
needed to determine the range of acceptable values in a given problem and,
beyond that, the results must be robust within this range for them to be
trustable. 

We thank A. K. Tagantsev for vivid and stimulating discussions. 

\end{document}